\documentclass[final,1p,11pt]{elsarticle}

\usepackage{epsfig}
\usepackage{array,tabularx,epsfig,mathrsfs,graphicx,rotating}
\usepackage{ifthen}
\usepackage{amsfonts}
\usepackage{ragged2e}
\PassOptionsToPackage{hyphens}{url}
\usepackage[hyphens]{url}
\usepackage{hyperref}
\usepackage{listings}
\usepackage{subfigure}
\usepackage{epstopdf}
\usepackage{color}
\usepackage{float}

\hypersetup{
  colorlinks=true,
  linkcolor=blue,
  citecolor=blue,
  urlcolor=blue
}

\graphicspath{{figs/}}

\newcommand{\beq}{\begin{equation}}
\newcommand{\eeq}{\end{equation}}

\chardef\til=126

\begin{document}

\definecolor{mygreen}{rgb}{0,0.6,0} \definecolor{mygray}{rgb}{0.5,0.5,0.5} \definecolor{mymauve}{rgb}{0.58,0,0.82}

\lstset{ %
 backgroundcolor=\color{white},   
 basicstyle=\footnotesize,        
 breakatwhitespace=false,         
 breaklines=true,                 
 captionpos=b,                    
 commentstyle=\color{mygreen},    
 deletekeywords={...},            
 escapeinside={\%*}{*)},          
 extendedchars=true,              
 keepspaces=true,                 
 frame=tb,
 keywordstyle=\color{blue},       
 language=Python,                 
 otherkeywords={*,...},            
 rulecolor=\color{black},         
 showspaces=false,                
 showstringspaces=false,          
 showtabs=false,                  
 stepnumber=2,                    
 stringstyle=\color{mymauve},     
 tabsize=2,                        
 title=\lstname,                   
 numberstyle=\footnotesize,
 basicstyle=\small,
 basewidth={0.5em,0.5em}
}

\begin{frontmatter}

\title{
Energy range of hadronic calorimeter towers and cells for high-$p_T$ jets at a 100~TeV collider
}

\author[add1]{S.V.~Chekanov\corref{cor1}}
\ead{chekanov@anl.gov}

\author[add1,add2]{J.~Dull}
\ead{dull@stolaf.edu}

\cortext[cor1]{Corresponding author}
\address[add1]{
HEP Division, Argonne National Laboratory,
9700 S.Cass Avenue,
Argonne, IL 60439, USA. 
}

\address[add2]{St. Olaf College,
1520 St. Olaf Avenue,
Northfield, MN, 55057, USA.
}


\begin{abstract}
This paper discusses a study of tower and cell energy ranges 
of a hadronic calorimeter for a $100$~TeV $pp$ collider.
The dynamic energy ranges were estimated using Standard Model jets with transverse
momenta above $20$~TeV.  
The simulations were performed using the PYTHIA Monte Carlo model  after 
a fast detector simulation tuned to the ATLAS hadronic calorimeter.
We estimate the maximum energy range of towers and cells as a function of lateral cell sizes for 
several extreme cases of jet transverse energy.
\end{abstract}

\begin{keyword}
hadronic calorimeter, jets, dynamic range, Monte Carlo, FCC
\end{keyword}

\end{frontmatter}


\section{Introduction}

The Future Circular Collider (FCC) is a proposed 80-100~km long ring that would collide protons at 100~TeV. 
A large experiment like this will require a detector that can measure high-energy particles and jets in the energy range of tens of TeV. 
For example, collisions at 100~TeV will produce hundreds\footnote{We assume the cross section of $0.02$ fb$^{-1}$ for the jet production with $p_T(jet)>20$~TeV and with an integrated luminosity of 10 ab$^{-1}$.} of Standard Model jets with transverse momentum $p_T(jet)>20$~TeV, while
many models beyond the Standard Model can lead to an even larger number of jets above $20$~TeV. 
This means that the calorimeter system of an FCC detector must be able to handle large energy depositions. 
The energy range of towers and cells of 
this future hadronic calorimeter must be well understood in order to 
set a stage for technology choices of future calorimeters.

The current calorimeters at the LHC are designed to measure jets with transverse momenta up to $4$~TeV.
The dynamic range of cells of a typical calorimeter at the LHC is about $10^4$, spanning the energy range of  $0.2-1500$~GeV 
for the ATLAS Tile calorimeter \cite{Aad:2010af,tang}. The lowest energy of 0.2~GeV is typically needed for muon reconstruction,
while the upper value of this range is set by high-$p_T$ jets.
It is clear that the dynamic range of a calorimeter which is designed to measure jets above $20$~TeV should be substantially extended.

In order to explore the maximum 
dynamic range of such a calorimeter, we use the HepSim repository \cite{Chekanov:2014fga} with Pythia8 predictions \cite{Sjostrand:2007gs},
and the DELPHES fast simulation \cite{deFavereau:2013fsa} after a proper tune of this simulation to describe energy sharing between different
layers of hadronic calorimeters.
We discuss the expected dynamic range for high-luminosity LHC (HL-LHC), and then we repeat this analysis for 100~TeV collision energies.
We assume that the calorimeter consists of electromagnetic and hadronic sections.
In our discussion, we also assume that a ``typical'' hadronic calorimeter has an interaction length ($\lambda$)  of 80\% of the
total interaction length of the entire calorimeter system.

\section{Fast simulation}

The fast detector simulation  uses the DELPHES framework which incorporates a 
tracking system, magnetic field, calorimeters, and a muon system \cite{deFavereau:2013fsa}.
DELPHES simulates the calorimeter system by summing together cells to form towers\footnote{Towers are regions defined in  pseudorapidity $\eta$ and 
azimuthal angle $\phi$.}. The towers are divided into electromagnetic and hadronic sections.
For the central analysis in this paper, we used towers of the size  $0.1 \times 0.1$ in pseudorapidity ($\eta$) and the azimuth angle ($\phi$).
Such tower sizes correspond to the sizes of the ATLAS hadronic calorimeter, the so-called Tile calorimeter.

The study of cell dynamic range requires tuning DELPHES to a full simulation of a hadronic calorimeter in order to reproduce
the longitudinal energy profile of hadronic showers. 
To tune DELPHES towers to the Tile calorimeter, 
we assume that all of the energies from photons, leptons and $\pi^0$ particles are deposited in the electromagnetic section.
Other particles deposit 60\% of their energy in the hadronic section \cite{Hernandez201383}, 
and 40\% in the electromagnetic section. 
It should be pointed out that these fractions are set to be constant and do not depend on particle's momentum, 
since DELPHES does not handle
energy sharing between electromagnetic and hadronic parts of a calorimter.

To convert towers to the cells we must go a step further. 
In the case of the ATLAS calorimeter,
about $50\pm 15\%$ of a jet's energy inside the hadronic calorimeter  is deposited to the first layer as follows
from the full detector
simulation  of  the ATLAS calorimeter system \cite{atlas_tile_cell}.
In order to convert tower energies to the cell energies, a scaling factor of 0.5 
was applied to the tower energies using  a Gaussian random distribution with the standard deviation 
of $0.15$ \cite{atlas_tile_cell}.

\section{Results}

\subsection{ATLAS-like calorimeter for HL-LHC}

\begin{figure}
\centering
  \subfigure[] {
  \includegraphics[scale=0.3]{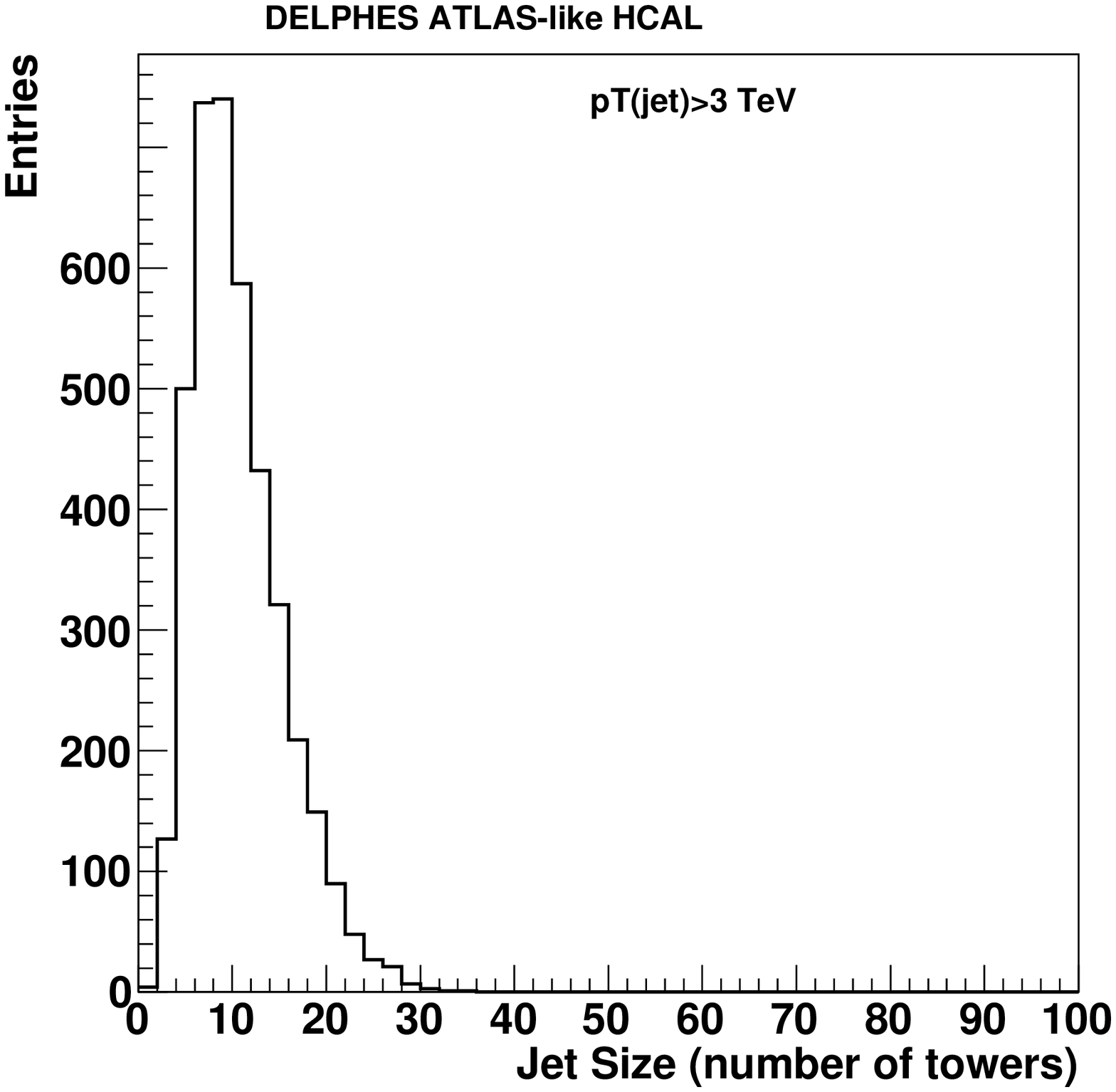}
  \label{fig:jetSize_1}}
  \subfigure[] {
  \includegraphics[scale=0.3]{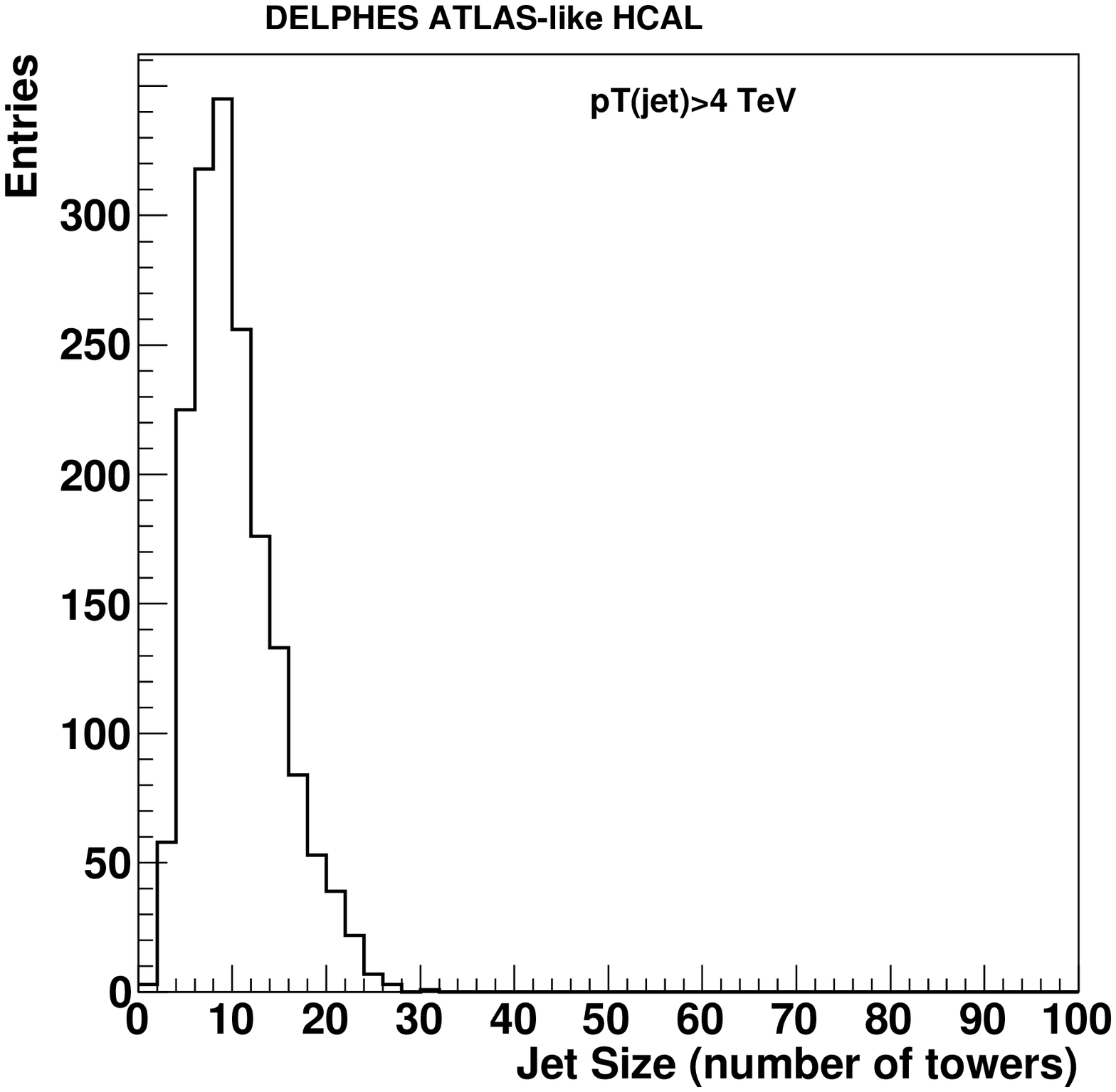}
  \label{fig:jetSize_2}}
\caption{The distributions of the  
numbers of towers of the size $0.1\times 0.1$ with energy above 0.2~GeV for jets at 
$p_T(jet)>3$~TeV (a) and $p_T(jet)>4$~TeV (b). 
The distributions were reconstructed 
using the anti-$k_t$ jets of the size 0.4 and the DELPHES ATLAS-like fast simulation.
The mean number of towers is 10.1 (9.7) for $p_T(jet)>3 (4)$~TeV.
} 
\label{fig:jetSize}
\end{figure}

\begin{figure}
\centering
  \subfigure[] {
  \includegraphics[scale=0.3]{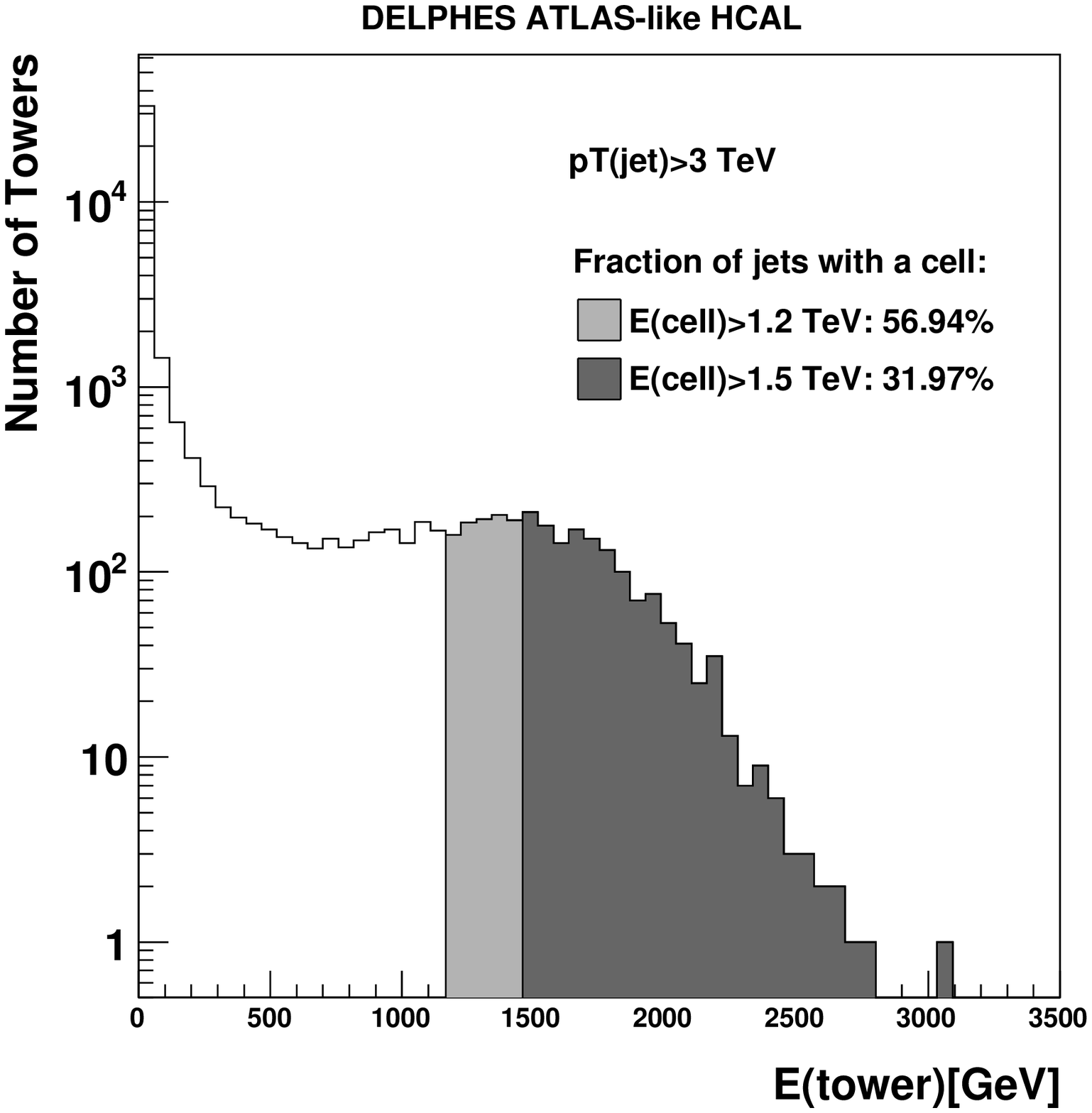}
  \label{fig:towersFull_1}}
  \subfigure[] {
  \includegraphics[scale=0.3]{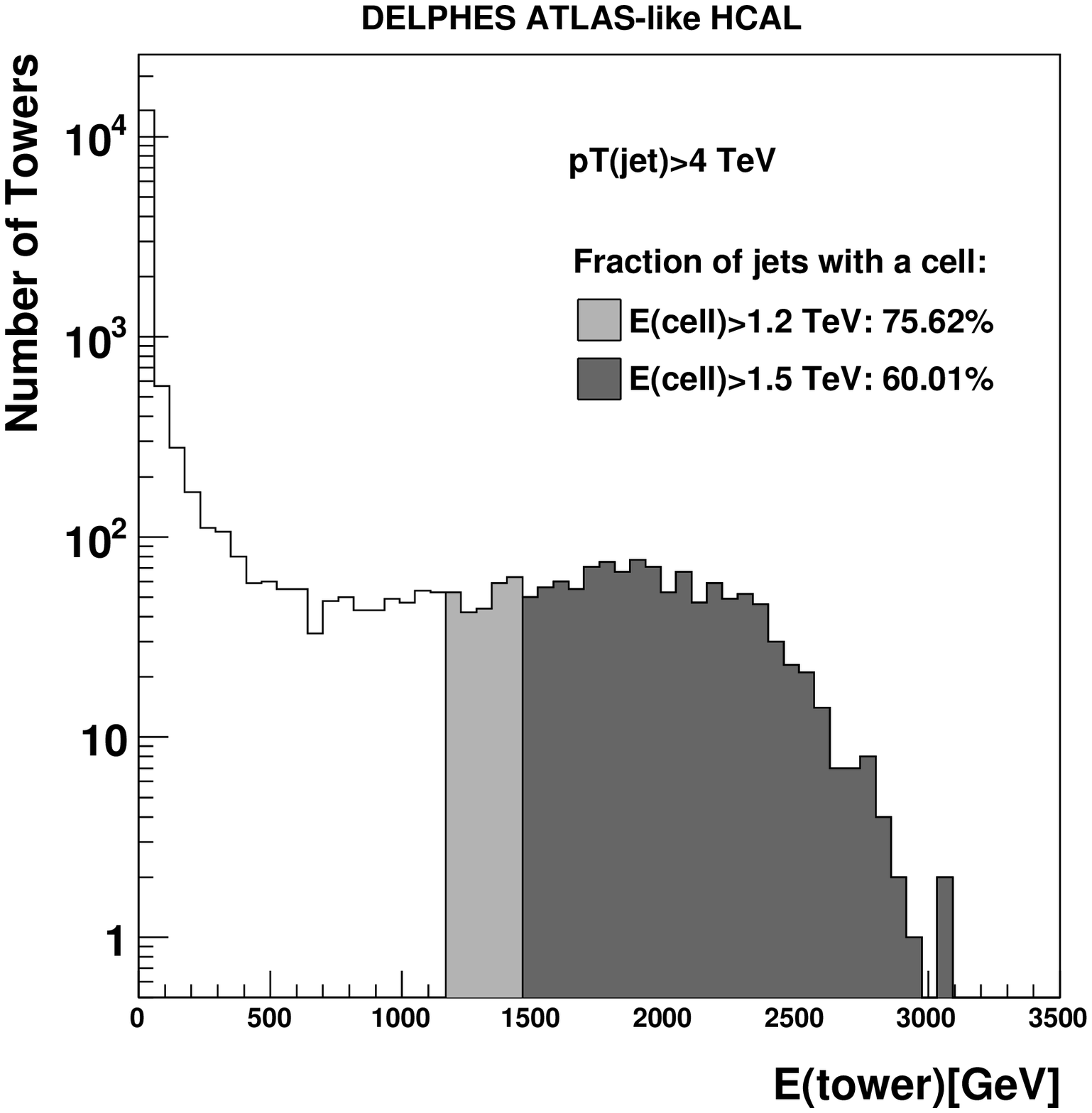}
  \label{fig:towersFull_2}}
\caption{The distributions of the energy of towers for jets with  $p_T(jet)>3$~TeV \subref{fig:towers_3tev} and $p_T(jet)>4$~TeV
\subref{fig:towers_4tev} assuming the size $0.1\times 0.1$ of the towers. }
\label{fig:DELPHES_Hcal_ATLAS_Full}
\end{figure}

\begin{figure}
\centering
  \subfigure[] {
  \includegraphics[scale=0.3]{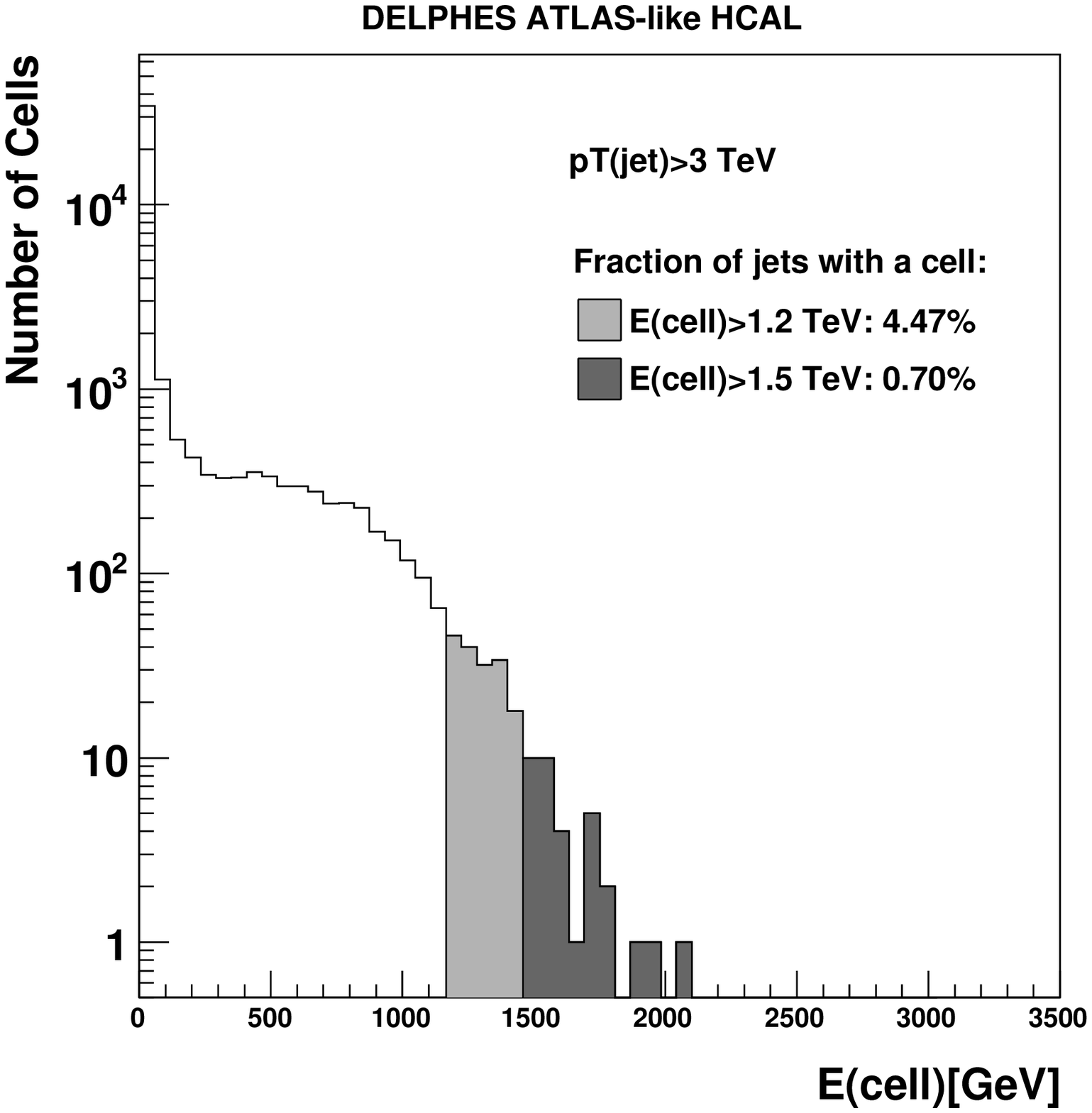}
  \label{fig:towers_3tev}}
  \subfigure[] {
  \includegraphics[scale=0.3]{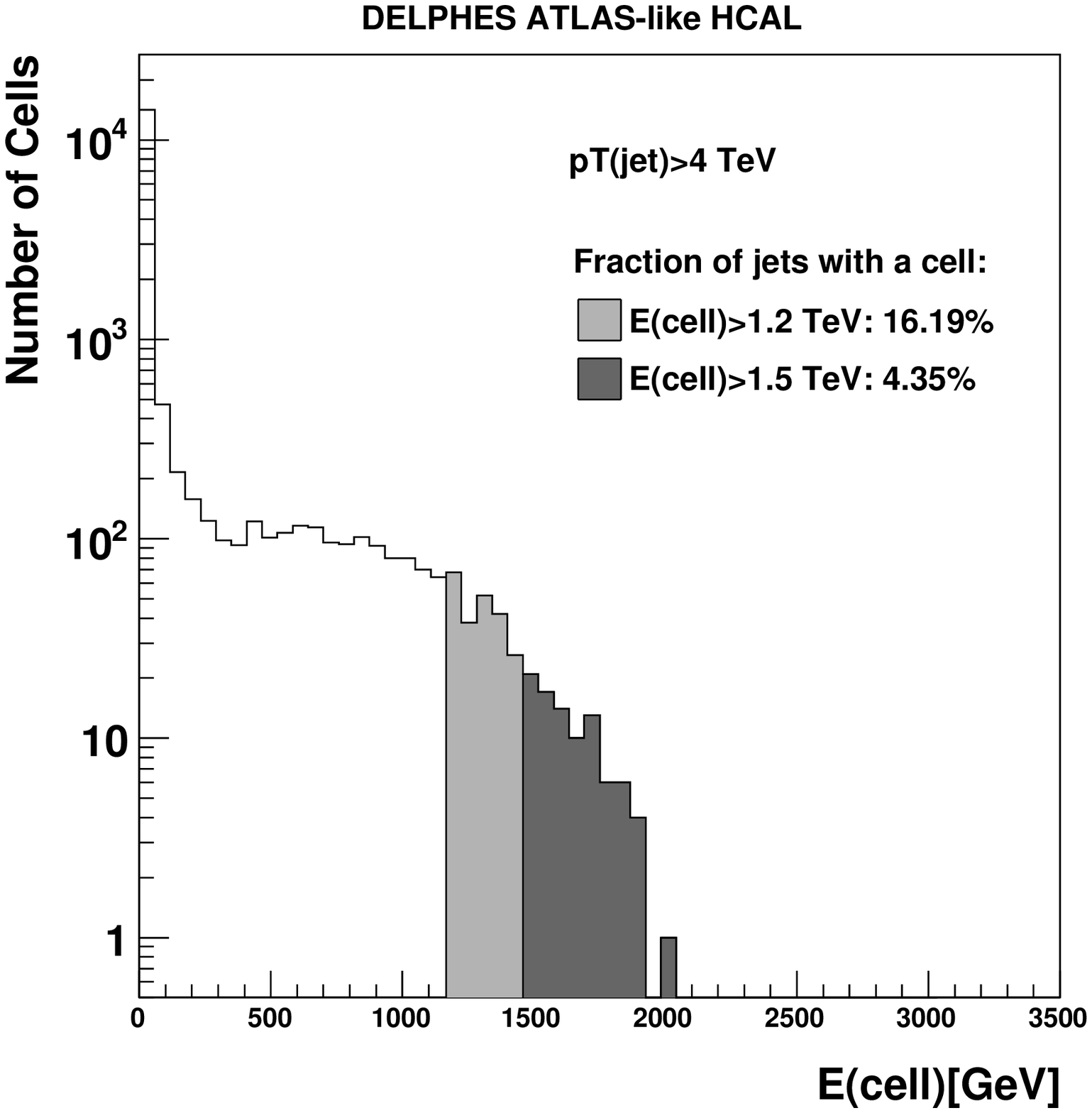}
  \label{fig:towers_4tev}}
\caption{The distributions of cell energy for jets with $p_T(jet)>3$~TeV \subref{fig:towers_3tev} and $p_T(jet)>4$~TeV \subref{fig:towers_4tev}. See other details in the text.}
\label{fig:DELPHES_Hcal_ATLAS}
\end{figure}

With DELPHES correctly tuned, Standard Model dijet events were generated by PYTHIA8 Monte Carlo model.
The particle-level event samples were downloaded from the HepSim repository \cite{Chekanov:2014fga}.
The jets are reconstructed using the anti-$k_T$ algorithm \cite{Cacciari:2008gp}
with  a distance parameter of 0.4 using the {\sc FastJet} package~\cite{fastjet}.
We select jets with $p_T(jet)>3$~TeV and $\eta<0.8$. 
In total, about 5,000 high-$p_T(jet)$ events were considered for this study.
Figure \ref{fig:jetSize} shows that approximately $10$ towers on average make up a $R=0.4$ jet in the DELPHES program.
This number does not change significantly with the transverse momentum cut in the range 3-4~TeV.  

If we consider whole towers to have a limited dynamic range of 1.2~TeV (1.5~TeV), then we find 70\% (32\%) of jets with at least one tower past this limit. 
For jets with $p_T(jet)>4$~TeV, these percentages rise to about 75\% and 60\% respectively. 
However, these fractions decrease when individual cells are simulated. Figure~\ref{fig:DELPHES_Hcal_ATLAS} shows 
that the fraction of jets affected by cell limits of 1.2 and 1.5~TeV drop to about 5\% and 1\%, respectively. 
Since multiple cells make up one tower, a single cell will never measure as much energy as a tower does. Thus, a cell is much less likely to saturate. 
The obtained limits are similar to those anticipated for the hadronic-calorimeter cells for the ATLAS phase II upgrade \cite{tang}. 

It should be noted that this simulation presents an optimistic view of these jet fractions. 
This study makes the assumption that  the  energy of hadrons in the hadronic part of a calorimeter system 
is fixed to $60\%$ as discussed
before. Note that DELPHES does not correct for energy-dependent effects of longitudinal propagation of particles.
Hence, the actual jet fractions may be higher than the values presented here.  
Nevertheless, the simulated numbers are close to those found by running a full simulation of the ATLAS detector.

\subsection{ATLAS-like calorimeter for 100 TeV collisions}

After the HL-LHC studies, we turn to jets that will be produced at a 100~TeV collider.
As before, we use the DELPHES simulation tuned to the ATLAS-like hadronic calorimeter
as described in the previous sections. The  dijets events  \cite{Chekanov:2014fga} were generated with PYTHIA8 
at a 100~TeV collision energy, and 
are reconstructed using the anti-$k_T$ algorithm discussed before.
For our studies, we use jets with $p_T(jet)>20$~TeV and $\eta<0.8$.

The lateral tower sizes of a calorimeter for a 100~TeV collider still need to be determined by looking at different physics cases.
One possible option is to reduce the tower size from  $0.1 \times 0.1$ to  $0.025 \times 0.025$.
With this option,
the energy in a single region of $0.1 \times 0.1$ will be shared by 16 towers with independent redouts.

\begin{figure}
\centering
  \subfigure[] {
  \includegraphics[scale=0.3]{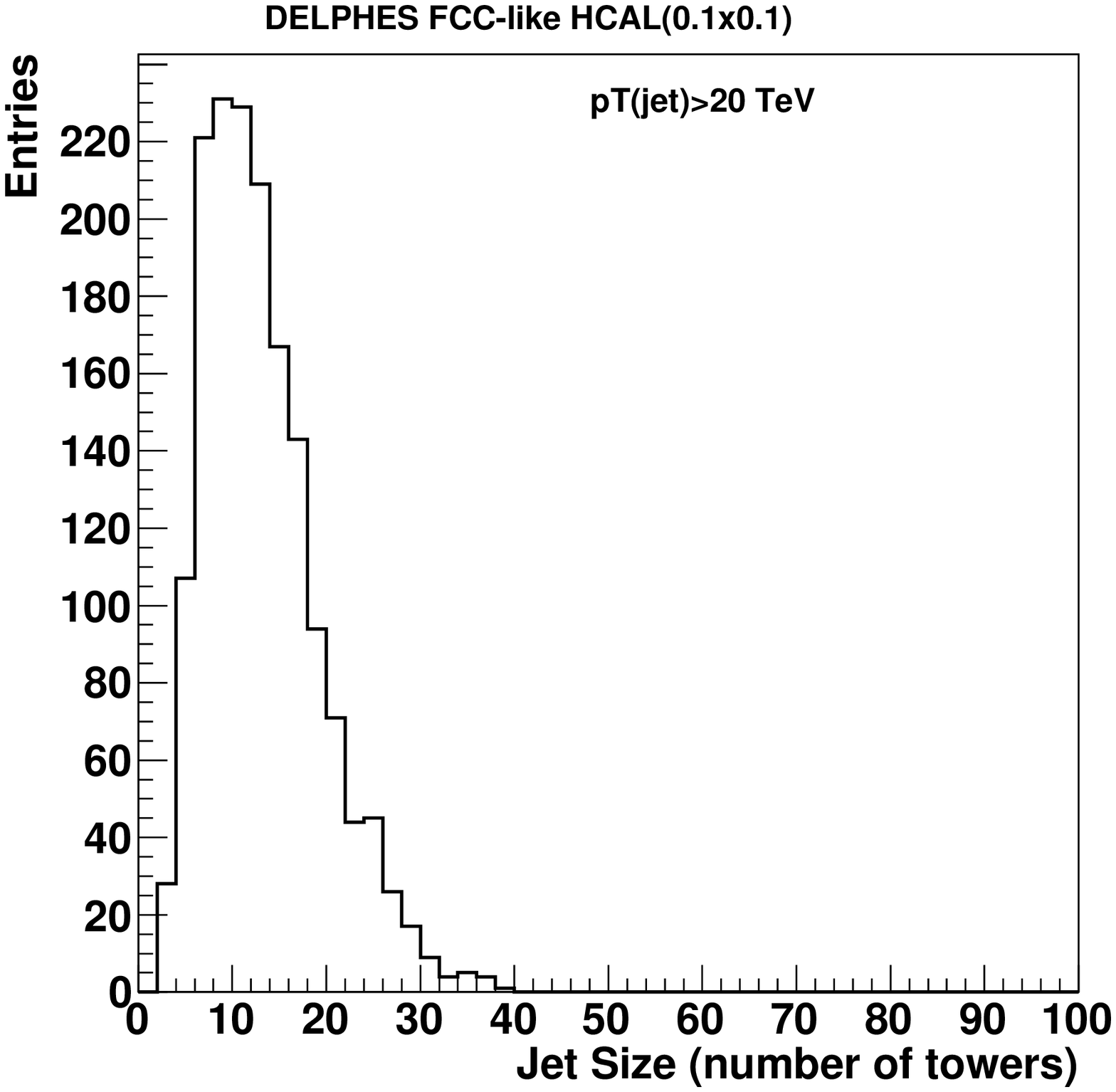}
  \label{fig:jetSize_4}}
  \subfigure[] {
  \includegraphics[scale=0.3]{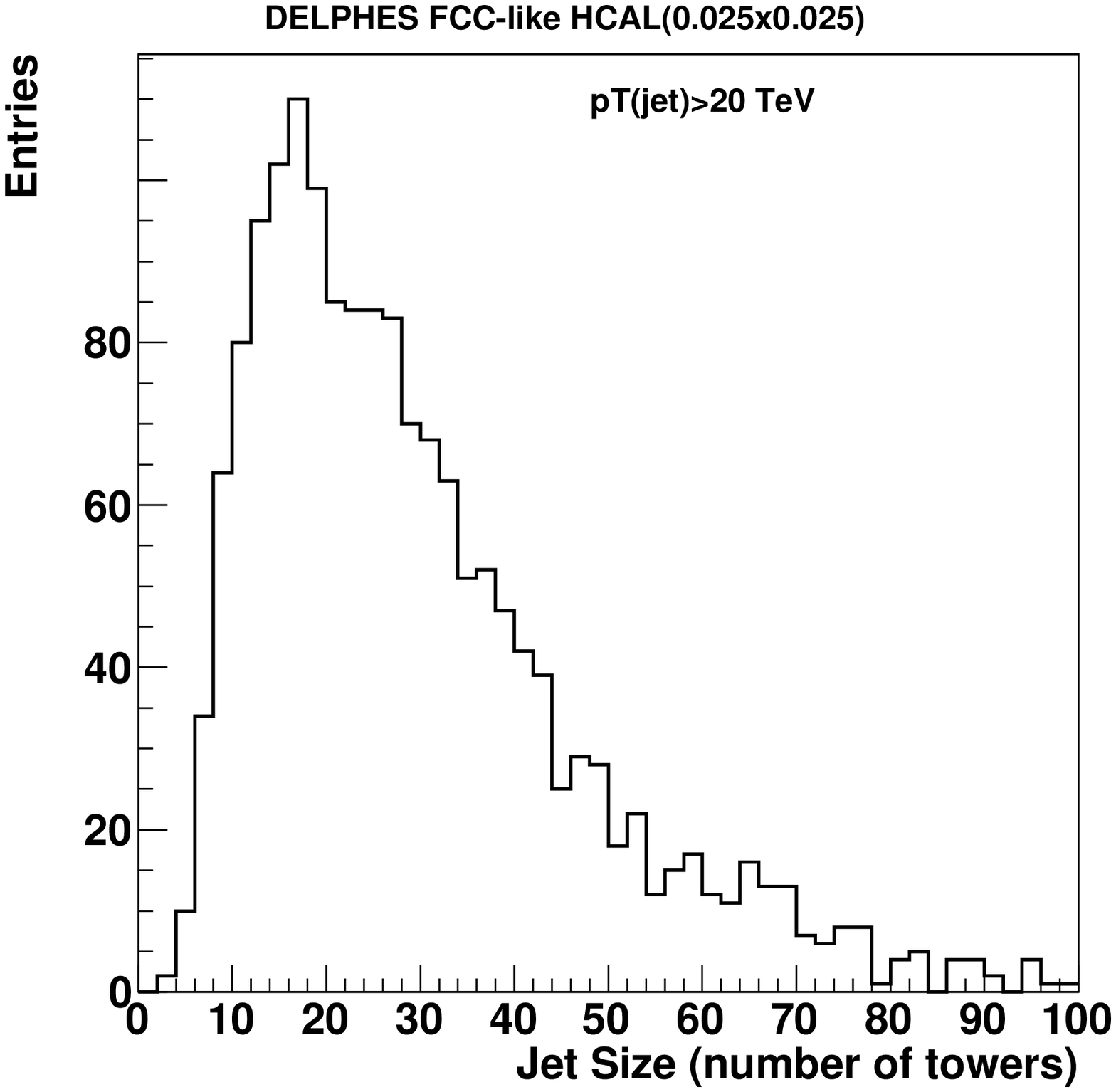}
  \label{fig:jetSize_3}}
\caption{The distribution of towers with 
energy above 0.2~GeV for jets with $p_T(jet)>20$~TeV and a $pp$ collision energy of 100~TeV. 
The average (RMS) values increase from  12.7 (6.1) for $0.1 \times 0.1$ \subref{fig:jetSize_4} to 28.9 (17.4) in the 
case with $0.025 \times 0.025$ \subref{fig:jetSize_3} towers.}
\label{fig:jetSize_FCC}
\end{figure}

The distribution of towers for jets with $p_T(jet)>20$~TeV measured in the FCC detector 
is shown in Figure~\ref{fig:jetSize_FCC}. For a tower division of $0.025 \times 0.025$, 
we find that the number of towers associated with jets increases by a factor two compared to a detector with a tower size of $0.1 \times 0.1$. 
If more towers make up a jet, we should expect that the energy range  for the towers will decrease.

\begin{figure}
\centering
  \subfigure[] {
  \includegraphics[scale=0.3]{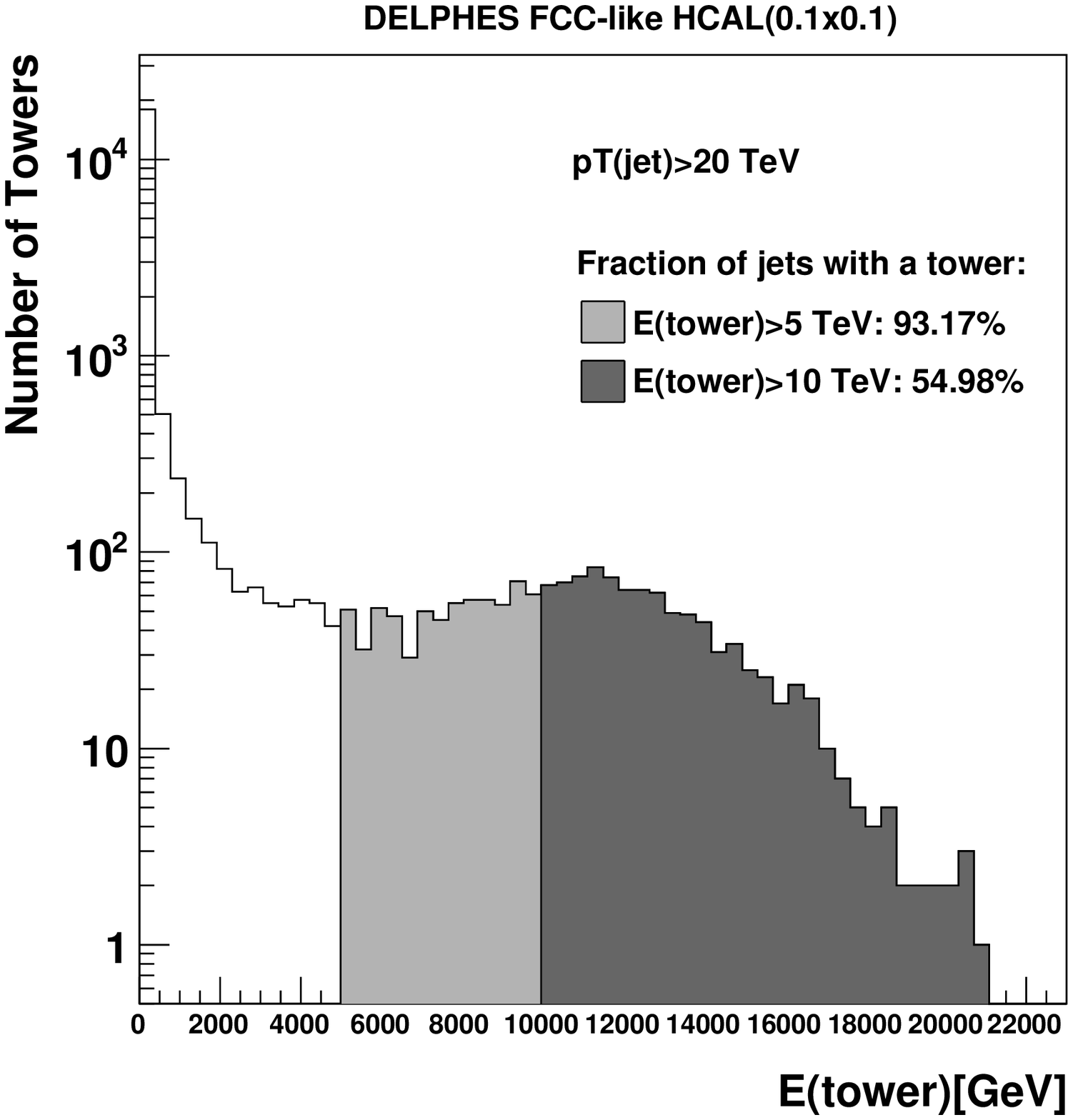}
  \label{fig:towersFull_4}}
  \subfigure[] {
  \includegraphics[scale=0.3]{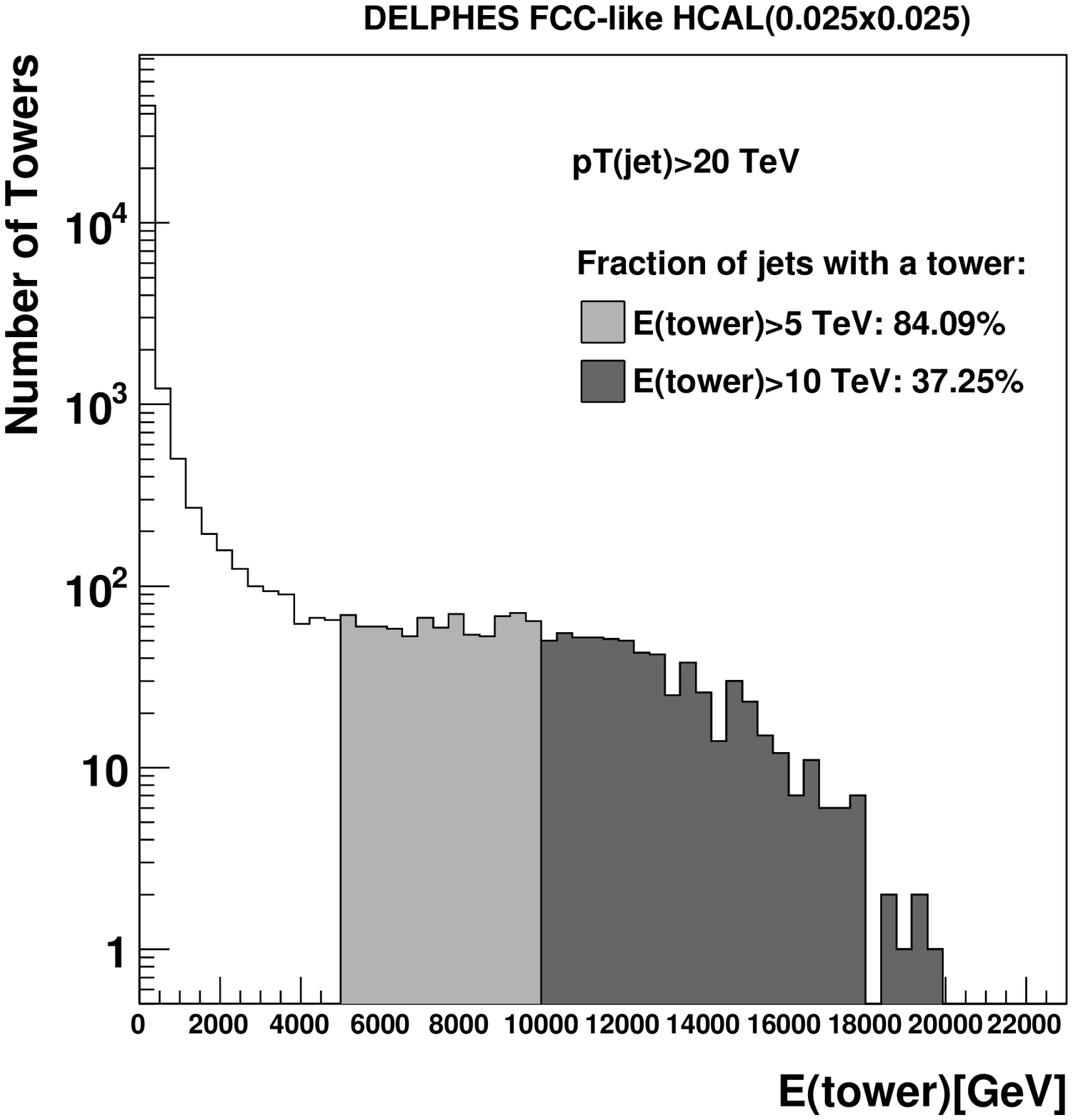}
  \label{fig:towersFull_3}}
\caption{The distribution of tower energy for ATLAS-like (left) and a FCC detector (right) 
detectors using cell sizes of 0.1 $\times$ 0.1 \subref{fig:towersFull_4} and  0.025 $\times $ 0.025 \subref{fig:towersFull_3}. 
We use $p_T(jet)> 20$~TeV jets.}
\label{fig:DELPHES_Hcal_FCC_Full}
\end{figure}

Figure~\ref{fig:towersFull_4} shows the distribution of energy seen by the towers of the size $0.1 \times 0.1$,
while Fig.~\ref{fig:towersFull_3} shows the same but for the towers of the size $0.025 \times 0.025$. 
The latter case indicates a smaller energy range. However, even for towers with the size $0.025 \times 0.025$,
there is a substantial amount of energy in the towers  beyond $10$~TeV.

\begin{figure}
\centering
  \subfigure[] {
  \includegraphics[scale=0.3]{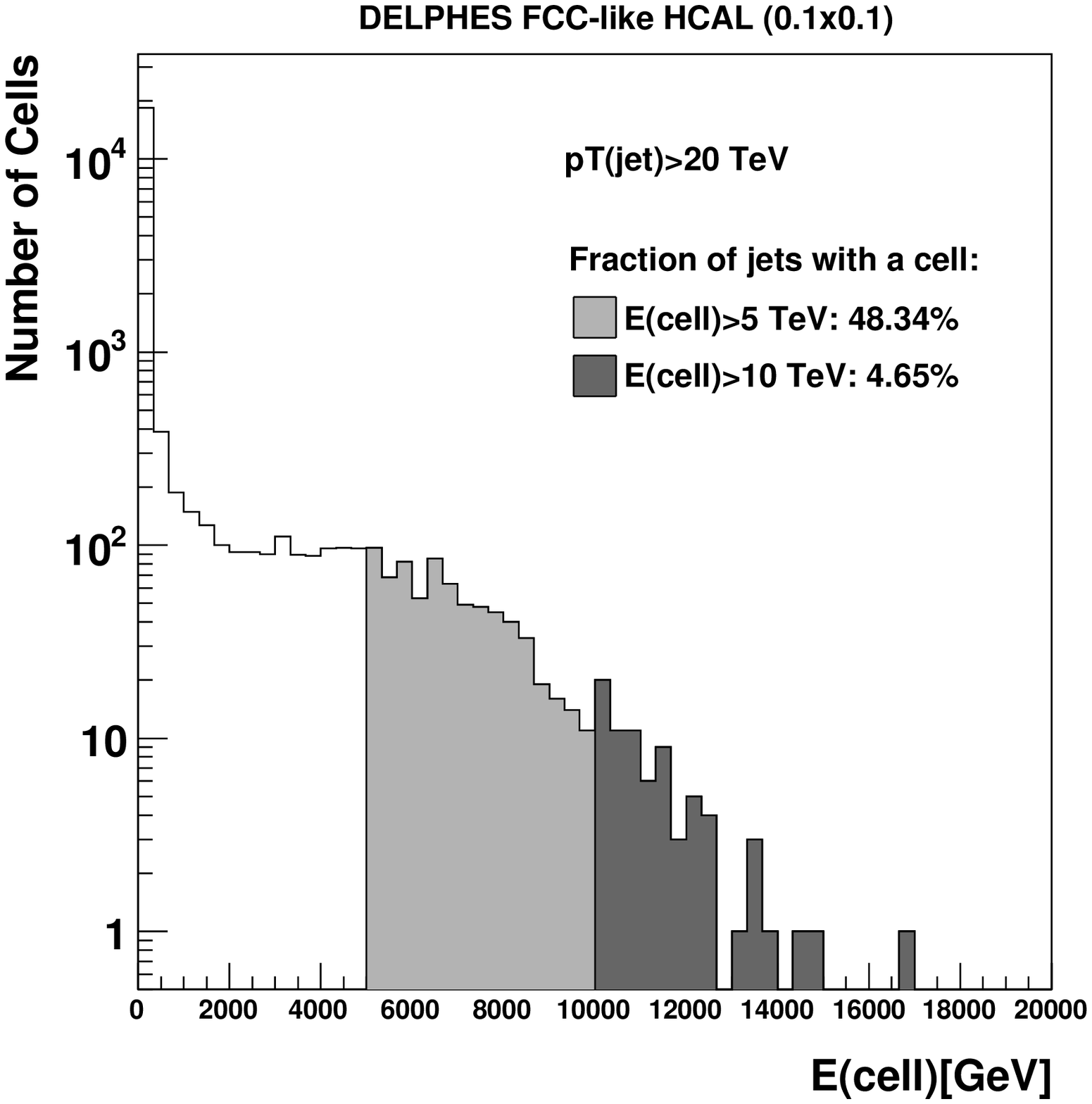}
  \label{fig:towers_fcc_atlas}}
  \subfigure[] {
  \includegraphics[scale=0.3]{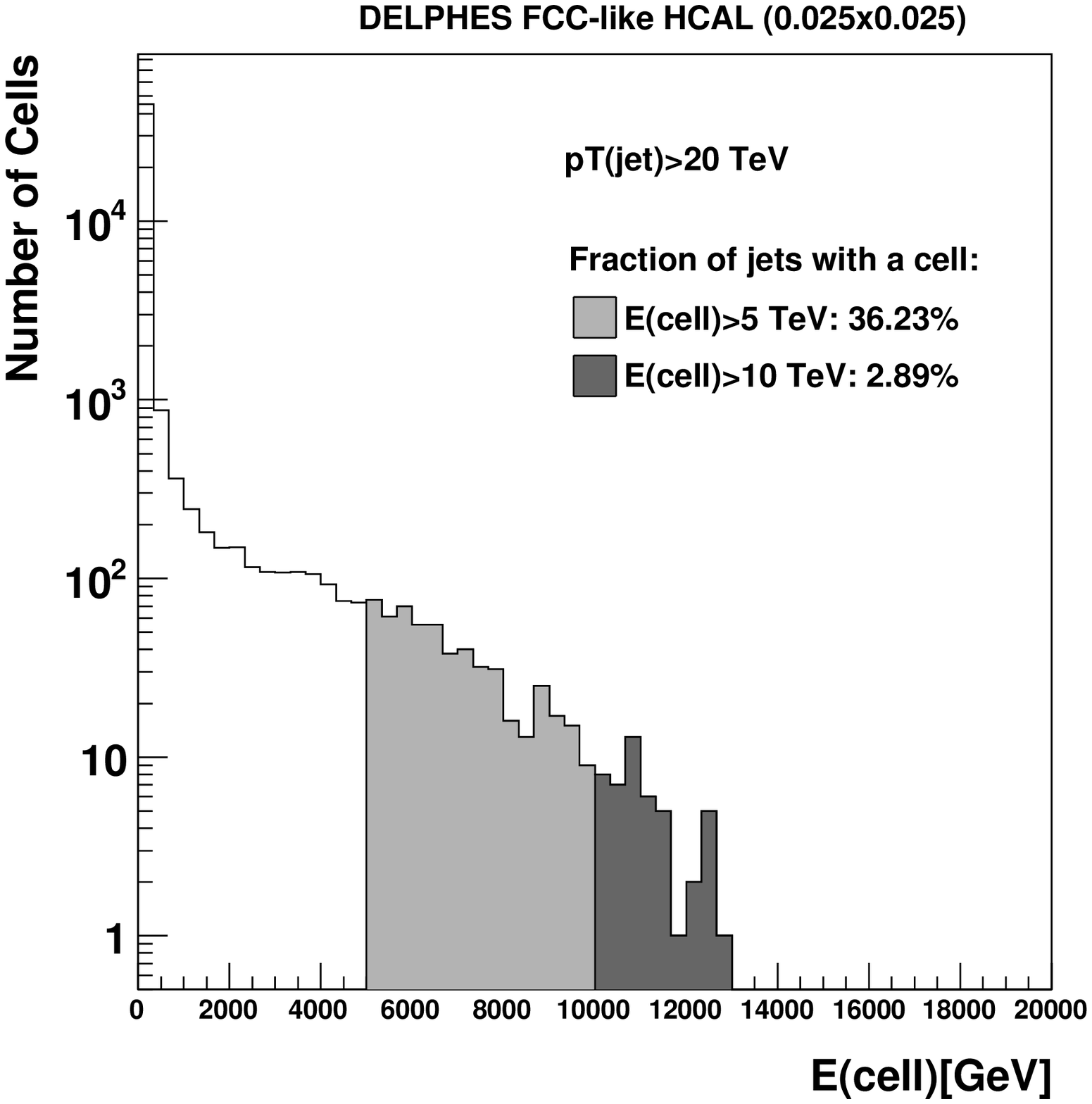}
  \label{fig:towers_fcc}}
\caption{The distributions of cell energy in TileCal for a FCC detector using cell sizes of 0.1 $\times$ 0.1 \subref{fig:towers_fcc_atlas} and  0.025 $\times $ 0.025 \subref{fig:towers_fcc}. Jets with  $p_T(jet)> 20$~TeV are used.}
\label{fig:DELPHES_Hcal_FCC}
\end{figure}

Figure~\ref{fig:DELPHES_Hcal_FCC} shows the energy distributions for calorimeter cells after
applying the tunning procedure from the Geant4 simulations as described in the previous section.
The fraction of jets with a cell above 5~TeV decreases from 48\% to 36\% when changing the cell size
from  $0.1 \times 0.1$ to $0.025 \times 0.025$. Figure~\ref{fig:towers_fcc_atlas} shows that nearly half of all 
high-$p_T$ jets have at least one cell of size $0.1 \times 0.1$ above 5~TeV.  
If cells are limited to the 10~TeV range, then only about 5\% of jets have a saturated cell reading. These numbers, 5 and 10~TeV, are used as our initial guesses to illustrate the maximum energy range for calorimeter cells anticipated for a 100 TeV collider.

We have calculated 99\% jet energy containment for anti-$k_T$ (R=0.4) jets  with $p_T(jet)>20$~TeV and 30~TeV.
It was defined as the maximum range needed for towers/cells 
to fully contain 99\% of the jet energy from the Standard Model processes. 
Assuming a tower size of $0.1\times 0.1$, the maximum energy containment in calorimeter towers is 18 (23) TeV   for $p_T(jet)>20(30)$~TeV.   
For cells of the same size, the energy containment values are 12 (16) TeV.
For $0.025\times 0.025$ tower (cell) sizes, the maximum energy containment decreases by roughly 20\% compared to the case with $0.1\times 0.1$ sizes. 
The size of this decrease was found to be smaller than naively expected, therefore, this observation needs to be verified with a full detector simulation.

\newpage
\section{Conclusion}
This paper discusses the energy range of calorimeter towers and cells for HL-LHC and for a future hadronic calorimeter at a 100~TeV collider.
We used a fast detector simulation in which
energy sharing between electromagnetic and hadronic parts are tuned to the ATLAS detector.

It was shown that about 1\% of jets with $p_T(jet)>3$~TeV will lose energy 
in reconstruction due to the present dynamic range of 0.2~GeV to 1.5~TeV of the ATLAS hadronic calorimeter. 
For 100~TeV $pp$ collisions, the dynamic range should be extended up to  18~TeV (towers) and 12~TeV (cells) for jets with $p_T(jet)>20$~TeV, which will be rather
common for 100~TeV collisions.  
This estimate assumes a similar lateral segmentation,
energy sharing between 
electromagnetic and hadronic calorimeters, and similar fraction of energy contributing to the first layer of hadronic calorimeter as for ATLAS.
These studies can be used to extrapolate the dynamic range values to smaller cell sizes.
It should be noted  that a provision for 30~TeV jets can be considered in the case of high-$p_T$ jet physics or 
exotic long-lived jets that deposit most of their energy in regions close
to the surface of the calorimeter. 

If the smallest energy to be reconstructed is 0.2~GeV as for the current designs
of hadronic calorimeters, the maximum range of 18~TeV may represent technological challenges for the readouts that should be able
to deal with $10^5$ dynamic energy range  (versus $10^4$ for the current calorimeters).

Despite the fact that the fast simulation was tuned to reproduce the full simulation of the ATLAS calorimeter, it should be noted that
the approximate nature of these calculations should fully be recognized when talking about 20-30 TeV jets. 
The quoted numbers  are likely to be on the optimistic side, since  
the energy fraction of hadrons in a hadronic calorimeter should be energy dependent. At present, this effect cannot be simulated with a fast simulation.
Nevertheless, we believe the obtained results can be used in early determination of the  design options for a future hadronic calorimeter, but should
later be checked when a full simulation becomes available.

\section*{Acknowledgements}
We thank J.~Proudfoot for the discussion. The submitted manuscript has been created by UChicago Argonne, 
LLC, Operator of Argonne National Laboratory (Argonne). Argonne, a U.S. Department of Energy Office of Science laboratory, is operated under Contract No. DE-AC02-06CH11357. Argonne National Laboratory's work was supported by the U.S. Department of Energy under contract DE-AC02-06CH11357. The U.S. Government retains for itself, and others acting on its behalf, a paid-up nonexclusive, irrevocable worldwide license in said article to reproduce, prepare derivative works, distribute copies to the public, and perform publicly and display publicly, by or on behalf of the Government.

\newpage

\section*{References}

\bibliographystyle{elsarticle-num}
\def\bibname{\Large\bf References}
\def\refname{\Large\bf References}
\pagestyle{plain}
\bibliography{biblio}

\end{document}